\documentclass{article}
\usepackage{graphicx}
\usepackage{amsmath}
\usepackage{subcaption}
\usepackage{algorithm2e}
\usepackage{setspace}
\usepackage{amssymb}
\usepackage{mathtools}
\usepackage{cancel}
\usepackage{algpseudocode}
\usepackage[export]{adjustbox}
\usepackage{blindtext}

\usepackage{comment}
\usepackage[colorlinks=true,linkcolor=blue,citecolor=blue]{hyperref}
\usepackage{geometry}
 \usepackage{nomencl}
 \usepackage{authblk}
\makenomenclature

\title{Trajectory Paradox is a Boundary Layer!}
\author[1]{Kush Kumar\thanks{E-mail: kushk20@iitk.ac.in,kushkumar408@gmail.com}}
\author[2]{Sovan Lal Das}
\author[1,3]{Shakti S. Gupta}
\affil[1]{Department of Mechanical Engineering, Indian Institute of Technology Kanpur, Kanpur, 208016, India}
\affil[2]{Physical and Chemical Biology Laboratory and Department
of Mechanical Engineering, Indian Institute of Technology Palakkad,
Palakkad, 678623, India}
\affil[3]{Department of Intelligent Systems, Indian Institute of Technology Kanpur, Kanpur, 208016, India}
\date{}

\begin{document}

\maketitle

\section*{Abstract}
This study derives the asymptotic conditions under which the massless stretched string formulation is valid for the equivalent dynamics of the inertial string carrying a moving point mass. Consequently, the governing equation under such conditions can be written as a singularly perturbed partial differential equation when the inertia of the string is included. When the inertia of the string is ignored, the governing equation reduces to a non-homogeneous hypergeometric ordinary differential equation governing the transverse motion of the moving mass. This hypergeometric equation has a discontinuous solution at the terminating boundary known as the trajectory paradox. However, numerically solving the actual equation yields a continuous solution with a rapid variation near the boundary. The complete solution resembles the inner solution as well, whereas the hypergeometric ordinary differential equation leads to an outer solution, revealing that the trajectory paradox is a boundary layer.
     
\section*{Keywords:}
Moving mass, Stretched string, Trajectory paradox, Galerkin method, Singularly perturbed differential equation, Green function 

\section{Introduction}
The analysis of moving loads on structures is a well-established area of research that remains highly relevant in the context of modern high-speed transportation. The interaction between a discrete moving load and a continuous structure gives rise to complex dynamic behavior, including transient vibrations, resonance effects, and significant structural displacements, making the analysis both mathematically intricate and practically important. 

The origins of moving load analysis can be traced to the pioneering work of Willis \cite{willis1849preliminary}, who investigated the dynamic behavior of beams subjected to moving loads. His study combined theoretical developments with experimental observations, allowing comparisons between structural responses under stationary and traveling loads. Soon afterward, Stokes \cite{stokes1849discussion} extended this line of research by deriving an exact analytical solution for an Euler--Bernoulli beam, simply supported at both ends and subjected to a moving concentrated mass. Unlike formulations that consider only the applied force, his model explicitly incorporated the inertia of the moving mass and examined a limiting case where the beam mass is negligible relative to the moving body. A noteworthy outcome of his analysis was that the moving mass would have a finite transverse displacement upon reaching the beam's end support, rather than satisfying the zero-displacement condition expected at the support. This unexpected result highlighted the subtle mathematical and physical complexities inherent in moving-mass problems and encouraged further investigations into their dynamic behavior.

Revisiting the moving mass problem nearly a century later, Smith \cite{smith1964motions} shifted the focus from simply supported beams to a horizontally stretched string. Consistent with Stokes's prior observation, Smith found that as the mass approaches the end of the string, its trajectory exhibits a discontinuity and fails to align with the expected fixed boundary. The author reduced the governing equation using the Green function, also given in \cite{fryba2013vibration}, to calculate the finite response at the terminating boundary. Dyniewicz and Bajer \cite{dyniewicz2009paradox} later explored this \emph{trajectory paradox} by applying a truncated Fourier sine integral series to estimate the transverse displacement, ultimately reporting similar behavior for an inertial string.

In recent developments, Gavrilov et. al. \cite{gavrilov2016revisitation} incorporated the string's wave pressure force acting on the moving mass, and also generalized the boundary condition that can accommodate a non-zero transverse displacement at the string's end to eliminate the paradox. Subsequently, Ferretti et al.\ \cite{ferretti2019nonlinear} developed the governing equations for a mass on a stretched string. By applying a Galerkin-based reduced-order model, their study successfully captured axial effects by allowing the tension to vary. A later study by Ferretti et al.\ \cite{ferretti2019dynamics} expanded on this approach by explicitly factoring in dynamic tension driven by axial deformation. In this work, they coined the term \emph{Stokes--Smith paradox} to describe the boundary anomaly and examined its convergence and the associated axial displacements. Most recently, Kumar and Gupta \cite{Kumardual} considered a dual-point contact moving mass on a string, which is a simple, moving rigid body rather than a point mass. They reported that the front contact trajectory exhibits a similar behavior to the trajectory paradox. As the two contact points of the moving mass approach their midpoint, the moving system reduces to an equivalent monopole-dipole system, as discussed in \cite{kumar2024response}.

In most of the work on trajectory discontinuity or the trajectory paradox, the authors considered a massless string without examining the conditions under which this is valid. Even if \cite{smith1964motions} stated a condition under which the massless string model is valid, but it is incomplete. None of the works has performed the asymptotic analysis required to relate the results obtained for a massless model to those for the corresponding exact inertial model. 

Therefore, in the present study, we begin by conducting an asymptotic analysis to determine the conditions under which the massless string model is valid. Corresponding to the asymptotic orderings, the governing equation of the system can be presented as a singularly perturbed equation. The unperterbed system (massless string) becomes the outer equation. Consequently, we show that the so-called trajectory paradox, or the boundary discontinuity of the outer solution, is not a true paradox, rather a standard boundary layer. Previous research has mischaracterized it as a paradox simply because the system was never treated as a singularly perturbed equation in the massless-string formulation.   

The remainder of this paper is organized as follows. Section \ref{sec:2} presents the governing equations of the system, along with the corresponding initial and boundary conditions. In Section \ref{sec:3}, we perform an asymptotic analysis of the non-dimensioned governing equation to derive the validity of the massless string model. Section \ref{sec:4} details the two solution strategies employed: the Galerkin method to solve the exact perturbed equation for the inner solution, and the Green's function approach to determine the outer solution. Section \ref{sec:5} presents results demonstrating the boundary-layer phenomena, and Section \ref{sec:6} offers concluding remarks.

\section{System and governing equation}{\label{sec:2}}

\begin{figure}[h]
    \centering     \includegraphics[width=0.7\linewidth]{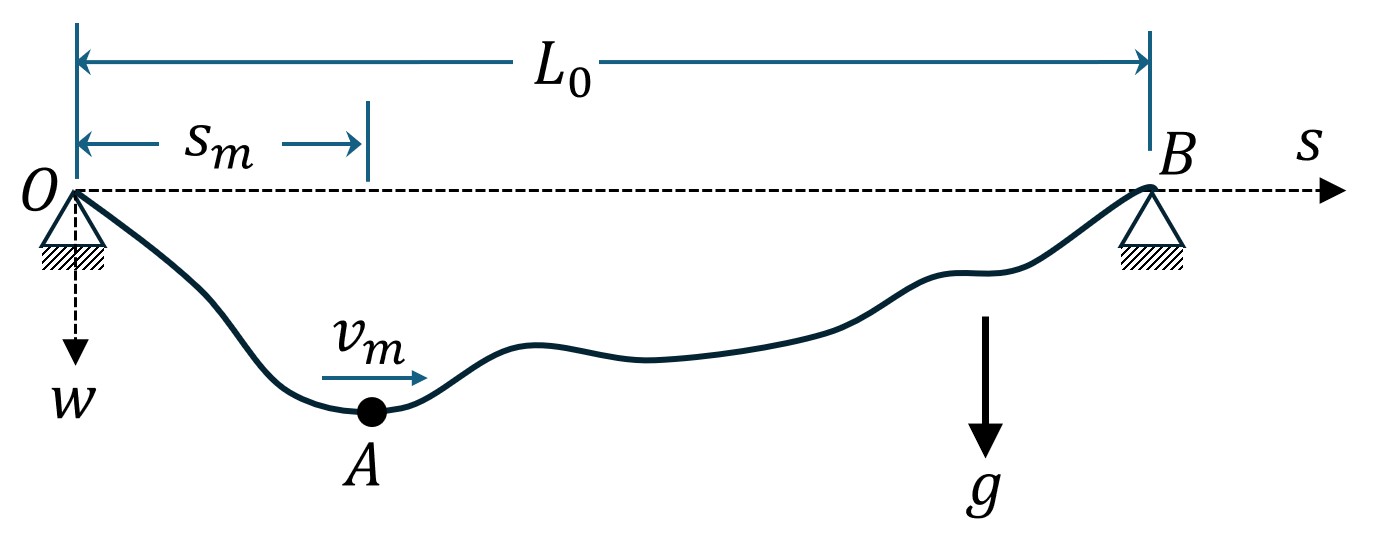}
    \caption{Linearized stretched string carrying a uniformly moving point mass.}
    \label{fig:1}
\end{figure}
The system of a linearized horizontal stretched string carrying a moving point mass is shown in Fig.~\ref{fig:1}. Line $OB$ is the initial state of the string, and curve $OAB$ is the current configuration of the string. The moving mass is currently located at $A$. It has magnitude $m$ and moves with a constant velocity $v_m$ along the length direction. $\rho_0$, $A_0$, and $L_0$ are the uniform mass density, uniform cross-sectional area, and the stretched length of the string. The tension in the string is $T$. It has been assumed that this tension remains constant during the mass's motion. Acceleration due to gravity $g$ acts downward (transverse) as shown in the figure.

The governing equation of the motion for this coupled system can be written as follows \cite{dyniewicz2009paradox}
\begin{equation}\label{eq:EOM_dim}
\begin{split}
    &\rho_0 A_0 w,_{tt}(s,t)-T w,_{ss}(s,t)=m\big(g- \Ddot{y}(t)\big)\delta(s-s_m),
\end{split}
\end{equation}
along with the initial and boundary conditions assumed as
\begin{equation}\label{eq:IC_BC}
    w(s,0) = \Dot{w}(s,t) = 0 \quad \text{and}\quad w(0,t) = w(L_0,t) = 0. 
\end{equation}
Here, $w(s,t)$ is the transverse displacement of a string's material point at longitudinal coordinate $s$ and at time $t$. For the string domain $s \in[0,L_0]$ and $s_m$ denotes the current prescribed longitudinal position of the moving mass, hence $v_m = \Dot{s}_m$. Using the condition of uniform velocity along the longitudinal direction, $s_m = s_m(t) = v_mt$, holds true. $\delta(s-s_m)$ is the Dirac delta function, which locates the point of interaction of the string with the moving point mass. Prescribed longitudinal displacement $(s_m)$ of the moving mass associated with its longitudinal motion, which starts when it is at the origin $O$, i.e., $s_m = 0$, and reaches the \emph{terminating boundary} $B$, i.e., $s_m = L_0$, at time $t = L_0/v_m$. $y(t)$ is the transverse displacement of the moving mass at time $t$. As it has been assumed that the mass is constrained to move on the string, the displacement compatibility relation can be written as follows
\begin{equation}\label{eq:dis_compatibility}
    y(t) = w\big(s_m(t),t\big).
\end{equation}
Finding the second total time derivative of the above equation, the transverse acceleration of the moving mass can be expressed as follows  
\begin{equation}\label{eq:accel_mass}
    \Ddot{y}(t) = \Ddot{w}\big(s_m(t),t\big)=\left.\left[v_m^2 w,_{ss}(s,t)+2v_m w,_{st}(s,t)+w,_{tt}(s,t)\right]\right|_{s=s_m(t)}.
\end{equation}

Throughout this work, $(\square),_{a}$ represents the partial derivative with respect to variable $a$. $(\Dot{\square})$ represents total derivative with respect to time $t$, $({\square}')$ represents total derivative with respect to position coordinate $s$, and $(\square|_{s = s_m})$ represents evaluation of an expression at $s = s_m$.

Using the relation given by Eq.~\eqref{eq:dis_compatibility} and the conditions given by Eq.~\eqref{eq:IC_BC}, the initial and final transverse conditions of the moving mass can be written as follows

\begin{equation}\label{eq:IC_FC}
    \begin{split}
        y(0) &= w(s_m(0),0) = 0\\ \Dot{y}(0) &= \Dot{w}\big(s_m(0),0\big) = v_m w,_s\big(s_m(0),0\big)+w,_t\big(s_m(0),0\big) = 0 \\ 
        y(0) &= w(s_m(0),0) = 0\\ y(L_0/v_m) &= w\big(s_m(L_0/v_m),L_0/v_m\big) = 0.
    \end{split}
\end{equation}
Initially, the transverse velocity of the moving mass is zero because the string is horizontally stretched and the initial slack is neglected. Consequently, the initial slope of the string is also zero, particularly at the origin, where the moving mass begins its motion. Furthermore, the first and the third conditions in the preceding relation are identical because the initial transverse displacement of the string at the fixed end $(s = 0)$ coincides with the corresponding boundary condition. Therefore, the initial and final conditions for the moving mass are given by
\begin{equation}\label{eq:IC_FC2}
    \begin{split}
        y(0) &= 0,\\ \Dot{y}(0) & = 0,\\ y(L_0/v_m) &= 0.
    \end{split}
\end{equation}

The first two relations in Eq.~\eqref{eq:IC_FC2} prescribe the initial conditions for the moving mass. In contrast, the third relation constitutes a final condition, which arises from the displacement compatibility condition in Eq.~\eqref{eq:dis_compatibility}, requiring the moving mass to remain in continuous contact with the string, together with the boundary condition of the string at $s=L_0$ given in Eq.~\eqref{eq:IC_BC}.

The transverse motion of the moving point mass is generally governed by the second-order ordinary differential equation
\begin{equation}\label{eq:EOM_mass}
    m\Ddot{y}(t)=mg-Q(t),
\end{equation}
where $Q(t)$ denotes the transverse reaction force exerted by the string on the moving mass at time $t$.

In the classical formulation, the motion of the point mass is uniquely determined by prescribing only its initial displacement and initial velocity. However, the third relation in Eq.~\eqref{eq:IC_FC2} additionally specifies the final position of the moving mass. Consequently, Eq.~\eqref{eq:EOM_mass} is supplemented by an additional constraint, thereby transforming the problem from a conventional initial-value problem into a constrained initial-value problem.

If Eq.~\eqref{eq:EOM_mass} is integrated using only the initial conditions (i.e., the first two conditions in Eq.~\eqref{eq:IC_FC2}),  under the massless string assumption, the resulting trajectory of the moving mass generally fails to satisfy the prescribed final condition near \(s_m=L_0\). This discrepancy is known as the \emph{trajectory paradox} in the classical massless-string model \cite{smith1964motions,dyniewicz2009paradox}. The inability of the reduced model to satisfy both the initial and final conditions suggests the existence of a \emph{boundary layer} near the terminal boundary, within which the solution adjusts to satisfy the prescribed final condition. Such behavior is characteristic of a singularly perturbed system. Motivated by this observation, in the following, we examine the asymptotic ordering for a massless string that leads the governing equation \eqref{eq:EOM_dim} to a singularly perturbed partial differential equation.

The following nondimensional parameters are introduced
\begin{equation}\label{eq:ND_para}
    \Bar{w} = \frac{w}{w_0}, \quad \Bar{y} = \frac{y}{w_0}, \quad \Bar{s} = \frac{s}{L_0}, \quad \Bar{s}_m = \frac{s_m}{L_0}, \quad \Bar{t} = \frac{t}{t_0},
\end{equation}
where $w_0 = mgL_0/(4T)$ is the characteristic transverse displacement of the static string, corresponding to its maximum deflection when the mass is attached at the midpoint, and $t_0 = L_0/v_m$ is the characteristic time taken by the moving mass to traverse the string length.

The governing equation in dimensionless quantities becomes
\begin{equation}\label{eq:EOM_ND2}
\begin{split}
    &\nu^2 \Bar{w},_{\Bar{t}\Bar{t}}(\Bar{s},\Bar{t})- \Bar{w},_{\Bar{s}\Bar{s}}(\Bar{s},\Bar{t})=\left(4- \frac{\nu^2}{\mu}\Ddot{\Bar{y}}\right)\delta(\Bar{s}-\Bar{s}_m),
\end{split}
\end{equation}
where $\nu = v_m/c$ is the ratio of the moving mass velocity to $c$, with $c = \sqrt{T/(\rho_0A_0)}$ denoting the wave speed in the string. The parameter $\mu = \rho_0A_0L_0/m$ represents the ratio of the string mass to the magnitude of the moving mass. Note that $\nu \ll 1$ when the mass is moving slowly and $\nu \sim 1$ when the mass is moving near the wave speed.
The initial and the final conditions of the moving mass from Eq.~\eqref{eq:IC_FC2} can be written in non-dimensional form as follows
\begin{equation}\label{eq:IC_FC_ND}
    \begin{split}
        \Bar{y}(0) &= 0,\\ \Dot{\Bar{y}}(0) & = 0,\\ \Bar{y}\big(L_0/(v_mt_0)\big) &= 0.
    \end{split}
\end{equation}

By prescribing suitable asymptotic orderings for $\nu$ and $\mu$, the governing equation can be reduced to a singularly perturbed differential equation with the inertial term of the string being only the perturbed term. In this reduction, the string's inertial effects become negligible, yielding a massless-string formulation that governs the dynamics of the moving mass.

\section{Asymptotic analysis}\label{sec:3}
First, let us consider only the case where $\nu \ll 1$. This is the case corresponding to the quasi-static motion of the system. That is when the mass moves slowly along the string. Also, note that from Eq. \eqref{eq:accel_mass} the non-dimensional transverse acceleration of the moving mass reduces to $\ddot{\bar{y}}$, which is $ \mathcal{O}(1)$. 

Now, to neglect the inertial effects of the string, corresponding to the massless string model defined in the literature \cite{stokes1849discussion,smith1964motions}, we assume $\nu^2 = \mathcal{O}(\epsilon)$, where $\epsilon$ is a small parameter. Under this assumption, Eq.~\eqref{eq:EOM_ND2} can be rewritten as
\begin{equation}\label{eq:case1_2}
\begin{split}
    &\epsilon\nu_0^2 \Bar{w},_{\Bar{t}\Bar{t}}(\Bar{s},\Bar{t})- \Bar{w},_{\Bar{s}\Bar{s}}(\Bar{s},\Bar{t})=\left[4- \frac{\epsilon\nu_0^2}{\mu}\Ddot{\Bar{y}}\right]\delta(\Bar{s}-\Bar{s}_m),
\end{split}
\end{equation}
where $\nu^2 = \epsilon \nu_0^2$ for $\nu_0^2 = \mathcal{O}(1)$. Ignoring the smaller terms from Eq.~\eqref{eq:case1_2}, the leading order equation can be written as follows 
\begin{equation}\label{eq:EOM_ND2_case1}
\begin{split}
    &- \Bar{w},_{\Bar{s}\Bar{s}}(\Bar{s},\Bar{t})=4\delta(\Bar{s}-\Bar{s}_m).
\end{split}
\end{equation}
The above equation demonstrates that, in the limit of a slowly moving mass, the inertial effects of both the string and the moving mass become negligible. As a result, the governing equation reduces to that describing a quasi-statically moving point force.

However, the inertial effects of the moving mass remain significant in the distinguished order for $\mu = \mathcal{O}(\epsilon)$, that is $\nu^2 \sim \mu$. In this case, Eq.~\eqref{eq:case1_2} takes the following form

\begin{equation}\label{eq:exact_massless}
\begin{split}
    &\epsilon\nu_0^2 \Bar{w},_{\Bar{t}\Bar{t}}(\Bar{s},\Bar{t})- \Bar{w},_{\Bar{s}\Bar{s}}(\Bar{s},\Bar{t})=\left[4- \frac{\nu_0^2}{\mu_0}\Ddot{\Bar{y}}\right]\delta(\Bar{s}-\Bar{s}_m),
\end{split}
\end{equation}
where $\nu^2 = \epsilon\nu^2_0$ and $\mu = \epsilon\mu_0$. Now, ignoring $\mathcal{O}(\epsilon)$,  Eq.\eqref{eq:exact_massless} reduces to
\begin{equation}\label{eq:EOM_ND3_case1}
\begin{split}
    &- \Bar{w},_{\Bar{s}\Bar{s}}(\Bar{s},\Bar{t})=\left[4- \frac{\nu_0^2}{\mu_0}\Ddot{\Bar{y}}\right]\delta(\Bar{s}-\Bar{s}_m).
\end{split}
\end{equation}
Therefore, the conditions $\nu^2 = \mathcal{O}(\epsilon)$ and $\mu = \mathcal{O}(\epsilon)$ represent the asymptotic ordering under which the massless string model remains valid for computing the response of a stretched string carrying a moving point mass. Furthermore, Eq.~\eqref{eq:exact_massless} is a singularly perturbed partial differential equation for $\nu^2 = \mathcal{O}(\epsilon)$ and $\mu = \mathcal{O}(\epsilon)$, since the highest derivative is multiplied by the small parameter $\epsilon$. Neglecting this small term yields the leading-order equation, Eq.~\eqref{eq:EOM_ND3_case1}.

It is worth noting that Smith \cite{smith1964motions} states that ``in order for the
massless-string approximation to be valid, the mass of the string must be small,
not in comparison with the mass of the moving particle, but in comparison with
the quantity $SL/v^2$.'' In our notation, $SL/v^2 = TL_0/v_m^2$. Therefore, the criterion proposed by the author becomes $\nu \ll 1$, or $v_m\ll c$.

This condition partially aligns with the asymptotic limit identified in the present study that leads to Eq.~\eqref{eq:EOM_ND3_case1}, corresponding to the quasi-static motion of the mass. Therefore, for the massless string model to remain valid, the conditions $v_m \ll c$ and $\rho_0A_0L_0 \ll m$ must be satisfied.

Note that no other asymptotic regime exists for which the massless string model remains meaningful. For example, taking $\nu=\mathcal{O}(1)$ and $\mu=\mathcal{O}(\epsilon)$ also leads to a massless string model; however, in this regime, the string exhibits reduced stiffness, while the gravitational force is asymptotically absent.

\section{Solution methods}\label{sec:4}

We solve equations Eq.~\eqref{eq:exact_massless} and Eq.~\eqref{eq:EOM_ND3_case1} using Galerkin and Green's function methods, respectively.

\subsection{Galerkin method for string with inertia}\label{sec:GM}
In the Galerkin method \cite{dyniewicz2009paradox,Kumardual} for  Eq.~\eqref{eq:exact_massless}, the transverse displacement of the string is approximated using the following truncated series

\begin{equation}\label{eq:GM_w}
    \bar{w}(\bar{s},\bar{t}) = \sum_{j=1}^{N} W_j(\bar{s}) p_j(\bar{t})
\end{equation}

where $W_j(\bar{s}) = \sin {j\pi \bar{s}}$ is a trial function that satisfies the non-dimensional form of the essential boundary condition given in Eq.~\eqref{eq:IC_BC} and $p_j(\bar{t})$ is the abstract modal coordinate of the $j$th trial function, and $N$ is the number of terms considered in the truncated series. Substituting $\bar{w}(\bar{s},\bar{t})$ from Eq.~\eqref{eq:GM_w} into Eq.~\eqref{eq:exact_massless} and taking the inner product with $W_i(\bar{s})$ over the non-dimensional length of the string $\bar{s} \in [0,1]$, we get

\begin{equation}
    \begin{split}
        &\frac{\epsilon \nu^2_0}{2} \Ddot{p}_i(\bar{t}) +\frac{\left(i\pi \right)^2}{2} p_i(\bar{t}) = 4\sin{\big(i \pi \bar{s}_m\big)}\\
        &-\frac{\nu^2_0}{\mu_0}\sin{\big(i\pi \bar{s}_m\big)}\sum_{j=1}^N \left(\sin{\big(j\pi \bar{s}_m\big)}\Ddot{p}_j(t)+2 j \pi \Dot{\bar{s}}_m\cos{\big(j\pi \bar{s}_m\big)}\Dot{p}_j(t)-\left(j \pi \Dot{\bar{s}}_m\right)^2\sin{\big(j\pi \bar{s}_m\big)}p_j(t)\right).
    \end{split}
\end{equation}

For $i=1,2,\ldots,N$, the above expressions constitute a system of $N$ discrete equations. In matrix form, this system can be written compactly as

\begin{equation}\label{eq:GM_dis}
    [\mathbf{M}]\{\Ddot{\mathbf{p}}\}+[\mathbf{G}]\{\Dot{\mathbf{p}}\}+[\mathbf{K}]\{\mathbf{p}\} = \{\mathbf{F}\},
\end{equation}
where the expression of the matrices and the force vector are given by
\begin{equation}
    \begin{split}
        \mathbf{M}_{ij} &= \frac{\epsilon \nu^2_0}{2} \delta_{ij}+\frac{\nu^2_0}{\mu_0}\sin{\big(i\pi \bar{s}_m\big)}\sin{\big(j\pi \bar{s}_m\big)},\\
        \mathbf{G}_{ij} &= 2 j \pi \Dot{\bar{s}}_m\frac{\nu^2_0}{\mu_0}\sin{\big(i\pi \bar{s}_m\big)}\cos{\big(j\pi \bar{s}_m\big)},\\
        \mathbf{K}_{ij} &= \frac{\left(j\pi\right)^2}{2} \delta_{ij}-\left(j \pi \Dot{\bar{s}}_m\right)^2\frac{\nu^2_0}{\mu_0}\sin{\big(i\pi \bar{s}_m\big)}\sin{\big(j\pi \bar{s}_m\big)},\\
        \mathbf{F}_i &= 4\sin{\big(i\pi \bar{s}_m\big)}\\
        \mathbf{p}_i &= p_i.
    \end{split}
\end{equation}
Here $\delta_{ij}$ is the Kronecker delta operator. Eq.~\eqref{eq:GM_dis} can be integrated to compute the system response for the non-dimensional form of initial conditions defined in Eq.~\eqref{eq:IC_BC}. The trajectory of the moving mass is $\bar{y}(\bar{t}) = \bar{w}(\bar{s}_m,\bar{t}) = \sum_{j=1}^{N} W_j(\bar{s}_m) p_j(\bar{t})$.

\subsection{Green function method for outer equation~\eqref{eq:EOM_ND3_case1}}\label{sec:GF}
The transverse displacement for the string in terms of the Green function is given by 

\begin{equation}\label{eq:GF_w}
    \bar{w}(\bar{s},\bar{t}) = \int_0^{1} \bar{G}(\bar{s},\bar{r})\bar{f}(\bar{r},\bar{t}) \, d\bar{r}.
\end{equation}
where the external force is (see Eq.~\eqref{eq:EOM_ND3_case1}) 
\begin{equation}
    \bar{f}(\bar{r},\bar{t}) = \left[4- \frac{\nu_0^2}{\mu_0}\Ddot{\Bar{y}}\right]\delta(\Bar{s}-\Bar{s}_m)
\end{equation}
and the Green's function is given by \cite{fryba2013vibration}
\begin{equation}
    \bar{G}(\bar{s},\bar{r}) = \begin{cases}
        (1-\bar{r})\bar{s}, &  0\leq \bar{s}\leq \bar{r}\\
        (1-\bar{s})\bar{r}, & \bar{r}< \bar{s} \leq 1.
    \end{cases}
\end{equation}
For the trajectory of the moving mass, substituting non-dimensional form of relation \eqref{eq:dis_compatibility} into Eq.~\eqref{eq:GF_w}, we get
\begin{equation}
    \begin{split}
        \bar{y}(\bar{t}) = \bar{w}(\bar{s}_m,\bar{t}) &= \int_0^{1} \bar{G}(\bar{s}_m,\bar{r}) \left[4- \frac{\nu_0^2}{\mu_0}\Ddot{\Bar{y}}\right]\delta(\bar{r}-\bar{s}_m) \, d\bar{r}\\
        &= \bar{G}(\bar{s}_m,\bar{s}_m) \left[4- \frac{\nu_0^2}{\mu_0}\Ddot{\Bar{y}}\right]\\
        & = (1-\bar{s}_m)\bar{s}_m\left[4- \frac{\nu_0^2}{\mu_0}\Ddot{\Bar{y}}\right]
    \end{split}
\end{equation}
Rearranging the above equation, it reduces to
\begin{equation}\label{eq:GF_EOM}
    \frac{\nu_0^2}{\mu_0}\bar{s}_m(1-\bar{s}_m)\ddot{\bar{y}}(\bar{t})+\bar{y}(\bar{t}) = 4\bar{s}_m(1-\bar{s}_m).
\end{equation}
The nondimensionalization gives $t_0 = L_0/v_m$ gives $\bar{s}_m = \bar{t}$ and Eq.~\eqref{eq:GF_EOM} reduces to 
\begin{equation}\label{eq:ND_GF}
    \frac{\nu_0^2}{\mu_0}\bar{t}(1-\bar{t})\ddot{\bar{y}}(\bar{t})+\bar{y}(\bar{t}) = 4\bar{t}(1-\bar{t}).
\end{equation}
This equation is a non-homogeneous hypergeometric ordinary differential equation with regular singular points at $\bar{t}=0$ and $\bar{t}=1$ \cite{smith1964motions}. To compute the trajectory (outer solution) of the moving mass, Eq.~\eqref{eq:ND_GF} is integrated subject to the first two initial conditions given by Eq.~\eqref{eq:IC_FC_ND}. However, since $\bar{t}=0$ is itself a regular singular point, imposing the initial conditions $\bar{y}(0)=0$ and $\Dot{\bar{y}}(0)=0$ causes the coefficient of the highest-order derivative in Eq.~\eqref{eq:ND_GF} to vanish, reducing the differential equation to an algebraic constraint at the initial point. Consequently, the resulting initial value problem is singular, and the prescribed initial conditions cannot be imposed directly on the reduced equation.

This singularity is applicable to the governing equation for the massless string model and does not arise in the exact equation. Nevertheless, the outer solution of Eq.~\eqref{eq:ND_GF} can be computed using a simple numerical regularization. Instead of initiating the integration at the singular point $\bar{t}=0$, the integration is started at a small positive time, for example $\bar{t}=10^{-6}$, with the shifted initial conditions
\[
\bar{y}(10^{-6})=0, \qquad
\Dot{\bar{y}}(10^{-6})=0.
\]
Since $\bar{t}=10^{-6}$ is very close to the origin, the resulting numerical solution provides an accurate approximation to the outer solution.

\section{Results and boundary layer}\label{sec:5}
Since the non-dimensional governing equation \eqref{eq:EOM_ND2} contains two non-dimensional parameters, we compute and analyze the solutions for the different parameter orderings discussed above.

Before presenting the boundary layer solution for the massless string model, we first examine the limiting case in which the inertia of the moving mass is also negligible. This corresponds to the quasi-static moving mass formulation governed by Eq.~\eqref{eq:EOM_ND2_case1}. The parameter is chosen as $\mu = 0.2 = \mathcal{O}(1)$, corresponding to a moving mass of magnitude $m = 0.2\,\rho_0A_0L_0$. The moving mass velocity is prescribed as $v_m = 0.01c$, yielding $\nu = 0.01$ and $\epsilon = 0.0001$ for $\nu_0 = 1$.

\begin{figure}[h]
    \centering     \includegraphics[width=0.8\linewidth]{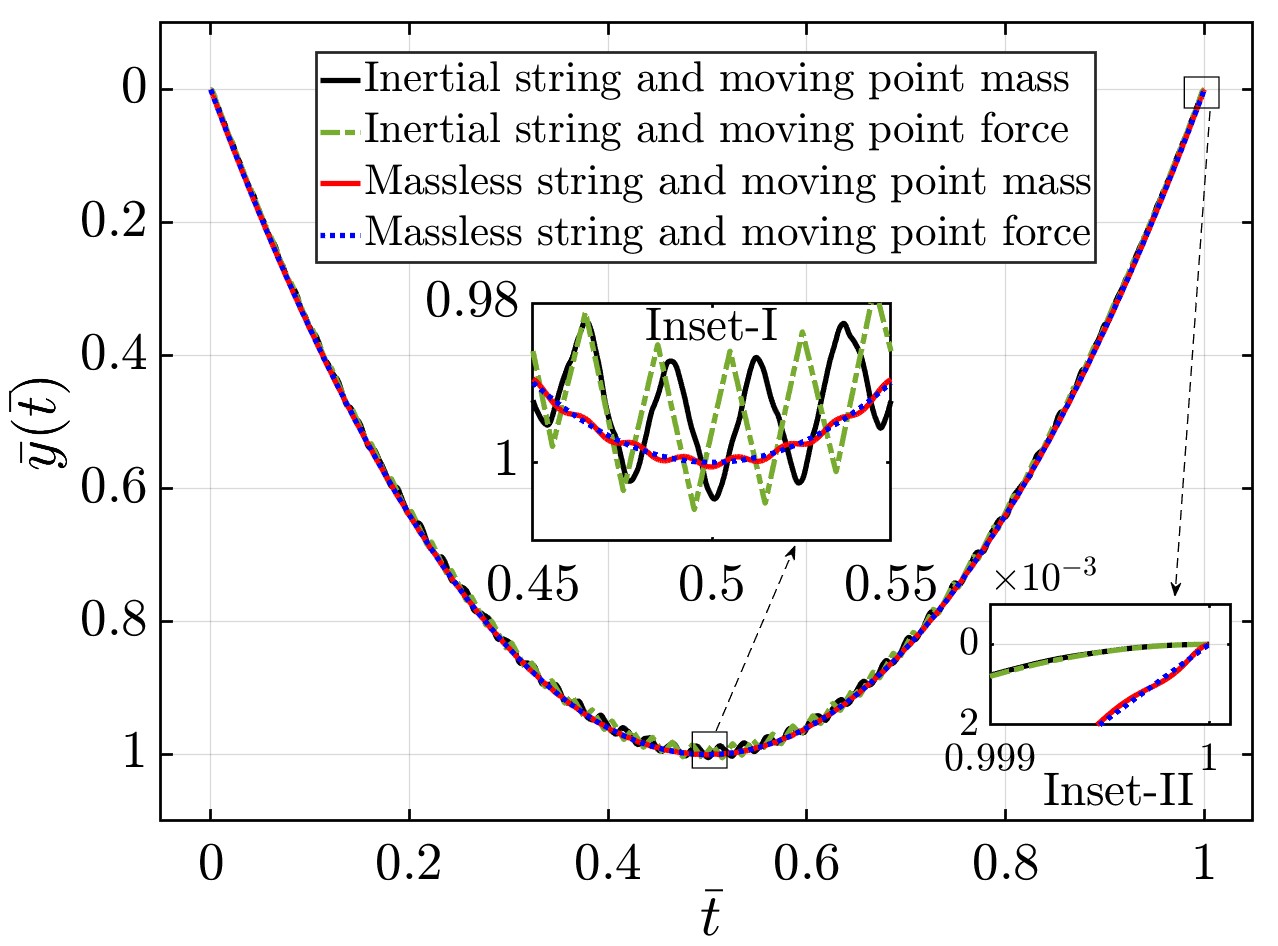}
    \caption{Trajectory responses of the moving mass for quasi-statically moving loads, computed for different cases.}
    \label{fig:r1}
\end{figure}

Fig.~\ref{fig:r1} shows the trajectory of the moving mass for the case $\nu=\mathcal{O}(\epsilon)$ and $\mu=\mathcal{O}(1)$. It is evident that all the computed trajectories are closely aligned. However, when the inertia of the string is retained, small local oscillations appear along the trajectory due to the propagation of high-speed wave disturbances in the string. These oscillations are observed for both the moving point mass model and its corresponding moving point force approximation, obtained by neglecting the inertia of the moving mass, as illustrated in Inset-I. Nevertheless, their amplitudes remain small compared to the overall transverse displacement.

The responses of the inertial string are computed using the Galerkin discretization described in Section~\ref{sec:GM}. In contrast, the dominant quasi-static response governed by Eq.~\eqref{eq:EOM_ND2_case1} is obtained using the Green's function formulation presented in Section~\ref{sec:GF}. Since this formulation neglects the inertia of the string, thereby representing a massless string, the wave-induced oscillations are eliminated. Consequently, the trajectories corresponding to the moving point mass and the associated moving point force become nearly indistinguishable for the massless string, as highlighted in the enlarged Inset-I.

Furthermore, from Eq.~\eqref{eq:ND_GF}, the trajectory of the moving point force on a massless string reduces to $\bar{y}(\bar{t}) = 4\bar{t}(1-\bar{t})$, which is a parabolic profile that aligns closely with plotted curves in Fig.~\ref{fig:r1}. When the inertial term in Eq.~\eqref{eq:ND_GF} is retained, its parametric coefficient $\epsilon\nu_0^2/\mu$ is sufficiently small, resulting in only weak oscillations about the parabolic trajectory $\bar{y}(\bar{t}) = 4\bar{t}(1-\bar{t})$. This behavior arises because the governing equation contains both inertial and stiffness terms, leading to small dynamic oscillations superimposed on the dominant quasi-static moving point force response.

Inset-II of Fig.~\ref{fig:r1} further shows that, for both the inertial and massless string models, the transverse displacement of the moving mass vanishes as it reaches the terminating support. Therefore, for the parameter regime $\nu=\mathcal{O}(\epsilon)$ and $\mu=\mathcal{O}(1)$, the massless string approximation correctly predicts zero transverse displacement at the terminating boundary. This implies that the final condition for the moving mass, i.e., the last condition in Eq.~\eqref{eq:IC_FC_ND}, is also satisfied, even though only the first two conditions of Eq.~\eqref{eq:IC_FC_ND} are imposed as the initial conditions for integrating Eq.~\eqref{eq:GF_EOM}.

Since the trajectory paradox is characterized by a non-zero transverse displacement of the moving mass at the terminating boundary for the massless string, thereby violating the final condition in Eq.~\eqref{eq:IC_FC_ND}, no such paradox arises under the present asymptotic ordering. To examine the emergence of the trajectory paradox, which we argue is a boundary-layer phenomenon, we next consider the parameter regime for which the governing equation reduces to Eq.~\eqref{eq:EOM_ND3_case1}.

\begin{figure}[h]
    \centering     \includegraphics[width=0.8\linewidth]{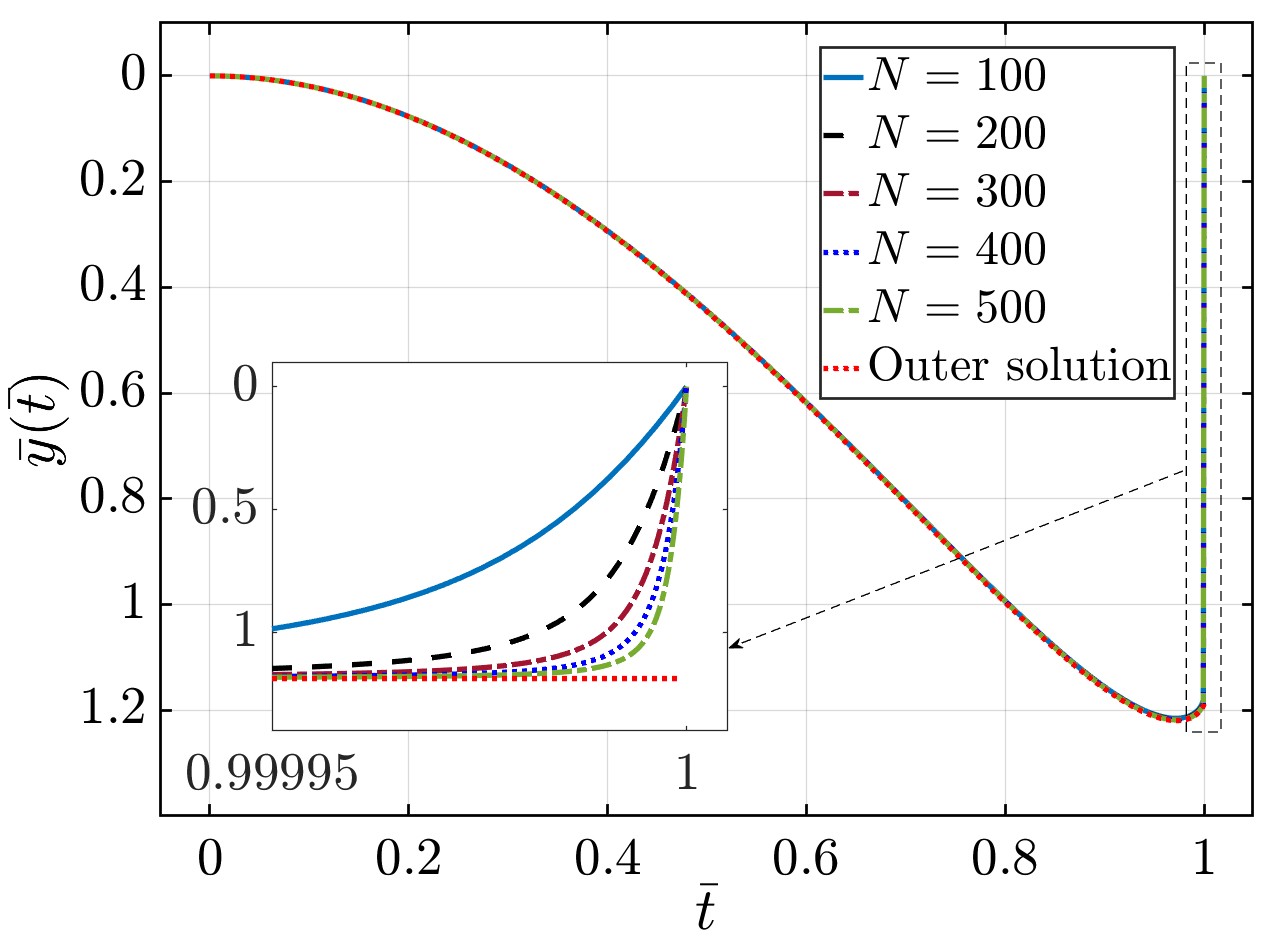}
    \caption{Inner and outer solutions of the singularly perturbed governing equation. The inner solution is computed for different values of $N$, which determine the truncation level of the Galerkin solution.}
    \label{fig:r2}
\end{figure}

Figure~\ref{fig:r2} illustrates the trajectories of the moving mass for the parameter values $\nu^2 = 0.01$ and $\mu = 0.01$. In dimensional terms, these correspond to a moving mass of magnitude $m = 100\,\rho_0 A_0 L_0$ traveling with velocity $v_m = 0.1c$. Introducing the small parameter $\epsilon = 0.01$, the dimensionless parameters can be expressed as $\nu^2 = \epsilon \nu_0^2$ and $\mu = \epsilon \mu_0$, where $\nu_0 = \mu_0 = 1$. The trajectories corresponding to the singularly perturbed governing equation, Eq.~\eqref{eq:exact_massless}, are computed using the Galerkin method (see Section~\ref{sec:GM}) and are presented for different truncation orders, $N$, of the Galerkin expansion. 

Setting $\epsilon = 0$ in Eq.~\eqref{eq:exact_massless} yields the reduced outer governing equation, Eq.~\eqref{eq:EOM_ND3_case1}. The corresponding solution is obtained through the Green's function formulation, leading to the non-dimensional Eq.~\eqref{eq:ND_GF}. Subsequently, integrating this equation subject to the first two conditions in Eq.~\eqref{eq:IC_FC_ND}, as described in Section~\ref{sec:GF}, which serve as the initial conditions for the moving mass, produces the trajectory shown as the outer solution in Fig.~\ref{fig:r2}.

As the moving mass approaches the terminating boundary, $\bar{s}_m = \bar{t} \rightarrow 1^{-}$, the outer solution exhibits a finite transverse displacement, as can be seen in the inset of Fig.~\ref{fig:r2}. Consequently, it fails to satisfy the final condition prescribed by the last equation in Eq.~\eqref{eq:IC_FC_ND}. This inconsistency is a characteristic feature of singular perturbation problems and provides clear evidence for the existence of a boundary layer in the trajectory of the moving mass.

The numerical solutions of the singularly perturbed governing equation further validate this behavior. As shown in the inset of Fig.~\ref{fig:r2}, increasing the truncation order of the Galerkin approximation progressively improves the solution near the terminating boundary. For a sufficiently large truncation order ($N=500$), the numerical solution accurately captures both the outer solution away from the boundary and the inner solution within the boundary layer. Away from the terminating boundary, the numerical trajectory is virtually indistinguishable from the solution of the reduced outer equation. Near the boundary, however, it develops the required boundary layer correction, enabling the trajectory to satisfy the final condition given by Eq.~\eqref{eq:IC_FC_ND}.

\begin{figure}[h]
    \centering     \includegraphics[width=0.8\linewidth]{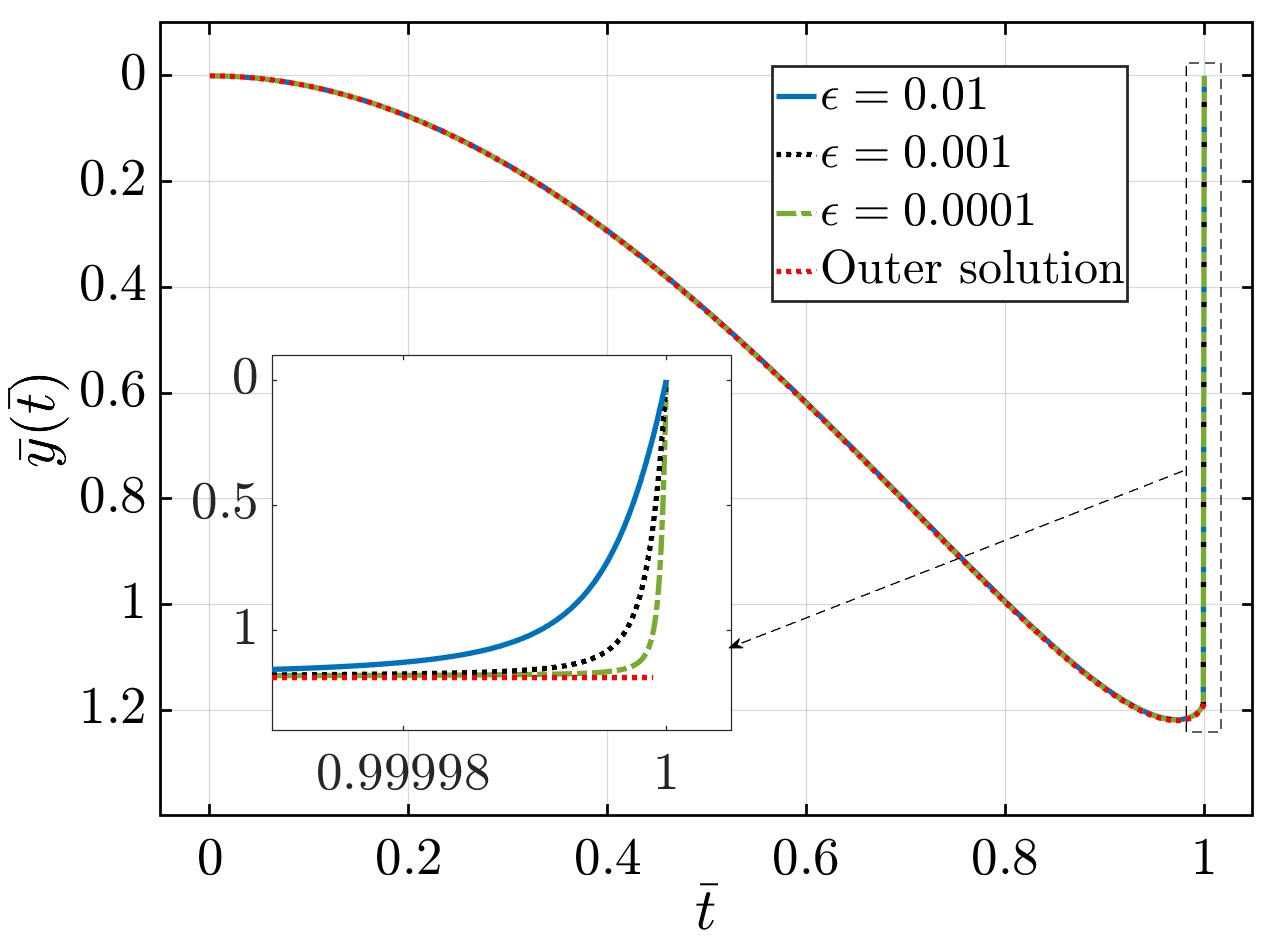}
    \caption{Inner and outer solutions of the singularly perturbed governing equation. The inner solution is obtained for various values of the perturbation parameter $\epsilon$.}
    \label{fig:r3}
\end{figure}

To further verify the boundary layer behavior, the singular perturbation parameter $\epsilon$ is progressively reduced, and Eq.~\eqref{eq:exact_massless} is integrated using the Galerkin formulation given by Eq.~\eqref{eq:GM_dis} with $N = 500$. The resulting trajectories for different values of $\epsilon$, together with the outer solution, which is independent of $\epsilon$, are presented in Fig.~\ref{fig:r3}. It is observed that, away from the terminating boundary, all trajectories obtained from the singularly perturbed governing equation coincide with the outer solution, confirming that the reduced outer equation accurately describes the solution in this region.

A closer examination of the terminating boundary, shown in the inset of Fig.~\ref{fig:r3}, reveals the development of distinct boundary layers at $\bar{t}=1$. As $\epsilon$ decreases, the width of the boundary layer becomes progressively smaller, while the numerical solution continues to follow the outer solution over an increasingly larger portion of the domain before rapidly transitioning within the boundary layer to satisfy the prescribed final condition.

Figure~\ref{fig:r2} shows that, as $N$ increases, the computed trajectories become progressively steeper in the inner (boundary layer) region. This indicates that $N$ should not be interpreted as a parameter analogous to the perturbation parameter $\epsilon$, since it does not determine the boundary layer thickness. Instead, increasing $N$ enriches the approximation by incorporating higher-order (higher-frequency) modes, enabling the numerical solution to resolve the rapid exponential variation within the boundary layer more accurately. Consequently, the steeper trajectories provide a better approximation to the boundary layer profile and allow the solution to satisfy the additional final boundary condition more closely. In contrast, when $N$ is fixed (e.g., $N=500$) and $\epsilon$ is varied, the expected boundary layer behaviour is recovered: decreasing $\epsilon$ reduces the boundary layer thickness while increasing the solution gradient in the inner region.

\section{Conclusion}\label{sec:6}
The perceived paradox regarding the trajectory of a moving point mass on a stretched string does not actually exist. The confusion stems from the massless string model, which is the dominant equation of a singularly perturbed differential equation. When this dominant equation is solved, it produces a trajectory discontinuity at the terminating boundary, which previous authors misinterpreted as a paradox. However, as demonstrated in this work, this is merely a boundary-layer phenomenon, with an inherent discontinuity of the dominant outer solution at the boundary.

\newpage
\bibliographystyle{elsarticle-num}
\bibliography{mybibfolder}
\end{document}